\documentstyle[12pt]{article}
\newcommand{\sect}[1]{\section{#1}\setcounter{equation}{0}}

\newcommand{\beq}{\begin{equation}}
\newcommand{\eeq}{\end{equation}}
\newcommand{\beqs}{\begin{eqnarray}}
\newcommand{\eeqs}{\end{eqnarray}}
\newcommand{\zm}[1]{\stackrel{o} {#1} }

\begin{document}
\begin{titlepage}
\null

\begin{flushright}
hep-th/9504026\\
KOBE-TH-95-01\\
April 1995
\end{flushright}

\vspace{7mm}
\begin{center}
  {\Large\bf Quantum Mechanics of Dynamical Zero Mode\par}
  {\Large\bf in $QCD_{1+1}$ on the Light-Cone\par}
  \vspace{1.5cm}
  \baselineskip=7mm

 {\large Motoi Tachibana\footnote{E-mail address:
tatibana@hetsun1.phys.kobe-u.ac.jp}\par}
\vspace{5mm}
  {\sl Graduate School of Science and Technology, Kobe University\\
     Rokkodai, Nada, Kobe 657, Japan \par}

\vspace{3cm}

{\large\bf Abstract}
\end{center}
\par

Motivated by the work of Kalloniatis, Pauli and Pinsky,
we consider the theory of light-cone quantized $QCD_{1+1}$
on a spatial circle with periodic and anti-periodic boundary conditions
on the gluon and quark fields respectively.This approach is based on
Discretized Light-Cone Quantization (DLCQ). We investigate the canonical
structures of the theory. We show that the traditional light-cone gauge
$A_{-}=0$ is not available and the zero mode (ZM) is a dynamical field,
which might
contribute to the vacuum structure nontrivially. We construct the full ground
state of the system and  obtain the Schr\"{o}dinger
equation for ZM in a certain approximation. The results obtained here
are compared to those of Kalloniatis et al. in a specific coupling region.
\end{titlepage}
\setcounter{footnote}{0}
\baselineskip=8.4mm


\sect{Introduction}

 Quantum field theory on the light-cone has been recently studied as a new
powerful tool for understanding non-perturbative phenomena \cite{pauli,horn},
especially
in the theory of strong interaction (QCD) \cite{brod,perry,kleba,wilson,zhang}.
One of the most remarkable advantages
in light-cone formalism is that vacuum is simple or trivial, i.e., Fock vacuum
is an eigenstate of the light-cone Hamiltonian \cite{harada}.
On the other hand, in the usual
equal-time formalism the vacuum contains an infinitely large number of
particles.
However, one simple and naive question arises here: how can we understand
phenomena like chiral symmetry breaking and confinement in such simple
vacuum ?

 As has already been indicated by many authors \cite{lenz,perry2,robert},
 zero modes of the fields might
play an essential and important role there. Recently Kalloniatis et al.
have investigated about pure glue $QCD_{1+1}$ (an SU(2) non-Abelian
gauge theory in 1+1 dimensions with classical sources coupled to the gluons)
and have discussed the physical effect of the dynamical zero mode \cite{kallo}.
Note here that there are two kinds of zero modes of the fields.
One is called constrained zero mode, which is not independent degrees of
freedom. Rather, it is dependent each other through the constraint equation.
There have been many works on such a constrained zero mode
related to the phenomena of  phase transition
in scalar field theory \cite{sriva,sande,maeno}.
The other is called dynamical zero mode we treat here, which is a true
dynamical
independent field. Also Kalloniatis et al. have used the specific approach of
Discretized Light-Cone Quantization (DLCQ) \cite{horn} in their analysis
because this
approach gives us an infrared regulated theory and the discretization of
momenta
facilitates putting the many-body problem on the computer. We shall follow
them, too.

 Our aim in this paper is to study the light-cone quantized $QCD_{1+1}$ with
fundamental fermions (quarks) coupled to the gauge fields (gluons) more in
detail
and give insight into the nontrivial QCD vacuum structure. The contents of
this paper are as follows. In Section 2, we study the canonical structures of
$QCD_{1+1}$ on the light-cone (Hamiltonian formalism) based on the Dirac's
treatment of the constraint system. We explicitly
obtain canonical light-cone Hamiltonian
and Dirac brackets between physical quantities there. We also comment on
dynamical
zero mode of the gluon fields in this section.
In Section 3, we quantize
the theory developed in the previous section and construct full ground state of
Hamiltonian . Furthermore we derive the Schr\"{o}dinger equation
for zero mode in a specific coupling region.
Section 4 is devoted to summary and discussion. The appendix is put in the
last part
of this paper for the explanations of notations and conventions.


\sect{Classical Theory  - Hamiltonian Formalism - }

  In this section, we study the canonical structures of light-cone
$QCD_{1+1}$, where the space is a circle and the gauge group is $SU(2)$.
 Let us start with following Lagrangian density
\begin{equation}
{\cal L} = -\frac{1}{4}F^{a}_{\mu \nu}F^{a\mu\nu} +
            \bar{\Psi}(i\gamma^{\mu}D_{\mu}-m)\Psi.
\label{lagrangian}
\end{equation}
Here $\Psi(x)$ is a quark field. Especially in two dimensions, the quark
field (in a representation in which $\gamma^5$ is diagonal)
\begin{equation}
\Psi(x) = \left( \begin{array}{c}
            \Psi_{R}(x) \\ \Psi_{L}(x)
            \end{array} \right).
\label{quark}
\end{equation}
is a two component spinor in the fundamental representation \cite{horn}.
R and L indicates chirality, which specifies only direction for massless
fermions. While the field $F^{a}_{\mu \nu}$ and the covariant derivative are
defined as
\begin{eqnarray}
F^{a}_{\mu \nu} &=& \partial_{\mu}A^{a}_{\nu}-\partial_{\nu}A^{a}_{\mu}
                  - g \epsilon^{abc}A^{b}_{\mu}A^{c}_{\nu},\\
iD_{\mu} &=& i\partial_{\mu} - gA^{a}_{\mu}T^a,
\label{fstreng,covderi}
\end{eqnarray}
where $A^{a}_{\mu}(x)$ is a \lq \lq gluon" field and $g$ is the coupling
constant
and $T^a$ and $\epsilon^{abc}$ are the generators and the structure constant
of the $SU(2)$ gauge group defined as  \cite{cheng}
\begin{eqnarray}
[ T^a, T^b ] &=& i\epsilon^{abc}T^c,\\
Tr(T^a T^b) &=& \frac{1}{2}\delta^{ab}.
\label{su(2)algebra}
\end{eqnarray}
In the light-cone frame approach, we set the coordinates
\begin{equation}
x^{\pm} = \frac{1}{\sqrt{2}}(x^0 \pm x^1),
\label{lccoordinate}
\end{equation}
and then rewrites all the quantities involved in the Lagrangian density
(\ref{lagrangian}) in terms of $x^{\pm}$ instead of the original coordinates
$x^0$(time) and $x^1$(space).
 As the usual Discretized Light-Cone Quantization, we define $x^{+}$ the
light-cone \lq \lq time", while $x^{-}$ the light-cone \lq \lq space",
which is restricted
to a finite interval from $-L$ to $L$. Within the interval, we impose
periodic and antiperiodic boundary conditions on the gluon field
$A^{a}_{\mu}(x)$ and the quark field $\Psi(x)$ respectively i.e.,
\begin{eqnarray}
A^{a}_{\mu}(x^+,x^{-}+2L) &=& A^{a}_{\mu}(x^{+},x^{-}),\\
\Psi(x^+,x^{-}+2L) &=&- \Psi(x^{+},x^{-}).
\label{bc}
\end{eqnarray}

In this way the Lagrangian density (\ref{lagrangian}) is rewritten as
\begin{eqnarray}
{\cal L} &=& \frac{1}{2}(F^{a}_{+-})^2\nonumber \\
         & &+\sqrt{2}(\Psi^{\dagger}_R i\partial_{+}\Psi_R
                       +\Psi^{\dagger}_L i\partial_{-}\Psi_L)
             -m(\Psi^{\dagger}_L\Psi_R + \Psi^{\dagger}_R\Psi_L)\nonumber \\
         & &-\sqrt{2}g(\Psi^{\dagger}_RT^{a}\Psi_R A^{a}_{+}
                       +\Psi^{\dagger}_LT^a\Psi_L A^{a}_{-}),
\label{lclagrangian}
\end{eqnarray}
where $\partial_{\pm}$ = $\frac{\partial}{\partial x^{\pm}}$.
In order to carry out the Hamiltonian formulation, we must compute the
canonical
momenta
\begin{eqnarray}
\Pi^{+a}(x) &=& \frac{\partial{\cal L}}{\partial(\partial_{+}A^{a}_{+}(x))}
                = 0,\\
\Pi^{-a}(x) &=& \frac{\partial{\cal L}}{\partial(\partial_{+}A^{a}_{-}(x))}
                = F^{a}_{+-}(x),\\
P_L(x) &=& \frac{\partial{\cal L}}{\partial(\partial_{+}\Psi_L(x))}
                = 0,\\
P^{\dagger}_L(x)
       &=& \frac{\partial{\cal L}}{\partial(\partial_{+}\Psi^{\dagger}_L(x))}
                = 0,\\
P_R(x) &=& \frac{\partial{\cal L}}{\partial(\partial_{+}\Psi_R(x))}
                = \sqrt{2}i\Psi^{\dagger}_R(x),\\
P^{\dagger}_R(x)
       &=& \frac{\partial{\cal L}}{\partial(\partial_{+}\Psi^{\dagger}_R(x))}
                = 0.
\label{momenta}
\end{eqnarray}

 On the other hand, we find that in the DLCQ approach, from the boundary
conditions (2.8), the gluon field $A^{a}_{\mu}$ can be decomposed into
zero mode (ZM) and particle mode (PM) as follows.
\begin{equation}
A^{a}_{\mu}(x) = \zm{A^{a}_{\mu}} + \widetilde{A}^{a}_{\mu}(x),
\label{decomp}
\end{equation}
where
\begin{eqnarray}
\zm{A^{a}_{\mu}} &\equiv& \frac{1}{2L}\int_{-L}^{L}dx^{-}A^{a}_{\mu}(x),\\
\widetilde{A}^{a}_{\mu}(x) &\equiv& A^{a}_{\mu}(x) - \zm{A^{a}_{\mu}}.
\label{zm}
\end{eqnarray}
Here $\zm{A^{a}_{\mu}}$ and $\widetilde{A}^{a}_{\mu}$ denote the zero modes
and the particle modes of the gluon field, respectively. Similarly the
canonical momenta (2.11) and (2.12) are decomposed into ZM and PM, which
leads to
\begin{eqnarray}
\zm{\Pi^{+a}} &=& 0,\\
\widetilde{\Pi}^{+a}(x) &=& 0,\\
\zm{\Pi^{-a}} &=& \zm{F}^{a}_{+-},\\
\widetilde{\Pi}^{-a}(x) &=& \widetilde{F}^{a}_{+-}(x),
\label{zmmom}
\end{eqnarray}
where
\begin{eqnarray}
\zm{F^{a}}_{+-} &\equiv&
\frac{1}{2L}\int_{-L}^{L}dx^{-}F^{a}_{+-}(x),\nonumber \\
                &=& \partial_{+}\zm{A^{a}_{-}}
                     -g\epsilon^{abc}\zm{A^{b}_{+}}\zm{A^{c}_{-}}
                    -\frac{g}{2L}\int_{-L}^{L}dx^{-}
\epsilon^{abc}\widetilde{A}^{b}_{+}(x)\widetilde{A}^{c}_{-}(x),\\
\widetilde{F}^{a}_{+-}(x) &\equiv& F^{a}_{+-}(x) - \zm{F^{a}}_{+-},\nonumber \\
                &=& \partial_{+}\widetilde{A}^{a}_{-}
                   -\partial_{-}\widetilde{A}^{a}_{+}
                  -g\epsilon^{abc}(\widetilde{A}^{b}_{+}\widetilde{A}^{c}_{-}
                   +\widetilde{A}^{b}_{+}\zm{A^{c}_{-}}
                   +\zm{A^{b}_{+}}\zm{A^{c}_{-}}).
\label{zmfstrength}
\end{eqnarray}
Following Dirac then \cite{dirac}, we can see
that there are six primary constraints such as \cite{sunder}
\begin{eqnarray}
& & \zm{\Pi}^{+a} \approx 0,\\
& & \widetilde{\Pi}^{+a}(x) \approx 0,\\
& & P_L(x) \approx  0,\\
& & P^{\dagger}_L(x) \approx 0, \\
& & P_R(x) - \sqrt{2}i\Psi^{\dagger}_R(x) \approx 0,\\
& & P^{\dagger}_R(x) \approx 0,
\label{primary}
\end{eqnarray}
where $\approx$ means the weak equality as usual. Substituting the equation
(\ref{decomp}) into the Lagrangian density (\ref{lclagrangian}) and using
the equations (2.11)-(2.16), total light-cone Hamiltonian can be obtained:
\begin{eqnarray}
H^{l.c.}_T &=& \int_{-L}^{L}dx^{-}[\Pi^{-a}\partial_{+}A^{a}_{-}
                      + P_R\partial_{+}\Psi_R-{\cal L}],\nonumber \\
          &=& \int_{-L}^{L}dx^{-}\left[\frac{1}{2}(\zm{\Pi}^{-a})^2
             +\frac{1}{2}(\widetilde{\Pi}^{-a})^2
           +\widetilde{\Pi}^{-a}\partial_{-}\widetilde{A}^{a}_{+} \right.
\nonumber \\
          & &+g\epsilon^{abc}(\zm{\Pi}^{-a}\zm{A^{b}_{+}}\zm{A^{c}_{-}}
+\widetilde{\Pi}^{-a}\widetilde{A}^{b}_{+}\widetilde{A}^{c}_{-}\nonumber \\
          & &+{\zm{\Pi}}^{-a}\widetilde{A}^{b}_{+}\widetilde{A}^{c}_{-}
             +\widetilde{\Pi}^{-a}\zm{A^{b}_{+}}\widetilde{A}^{c}_{-}
+\widetilde{\Pi}^{-a}\widetilde{A}^{b}_{+}\zm{A^{c}_{-}})\nonumber \\
          & &-\sqrt{2}\Psi^{\dagger}_Li\partial_{-}\Psi_L
             +m(\Psi^{\dagger}_L\Psi_R + \Psi^{\dagger}_R\Psi_L)\nonumber \\
          & &  \sqrt{2}g\left(\Psi^{\dagger}_RT^{a}\Psi_R
                             (\zm{A^{a}_{+}}+\widetilde{A}^{a}_{+})
                        + \Psi^{\dagger}_LT^{a}\Psi_L
(\zm{A^{a}_{-}}+\widetilde{A}^{a}_{-})\right)\nonumber \\
          & &+ u_0^{a}\zm{\Pi}^{+a} + u_1^{a}\widetilde{\Pi}^{+a} + u_2P_L
             +u_3P_L^{\dagger}\nonumber \\
          & & \left.+u_4(P_R-\sqrt{2}i\Psi_R^{\dagger}) +u_5P_R^{\dagger}
\right],
\label{lchamiltonian1}
\end{eqnarray}
where $u_0^{a}$ and $u_1^{a}$ ($u_2,u_3,u_4$ and $u_5$) are (Grassmann)
Lagrange multipliers and we have used the following facts
\begin{equation}
\int_{-L}^{L}dx^{-}\widetilde{A}^{a}_{\pm}(x)
  = \int_{-L}^{L}dx^{-}\widetilde{\Pi}^{\pm a}(x) = 0.
\label{pmcond}
\end{equation}

 Once we obtain the expression for the Hamiltonian, we must investigate
whether the primary constraints induce the secondary constraints by imposing
the consistency conditions.
 As the result, we find there are four secondary constraints:
\begin{eqnarray}
& & g\int_{-L}^{L}dx^{-}(\Pi^{-c}\epsilon^{abc}A^{b}_{-}+
                     \sqrt{2}\Psi^{\dagger}_RT^{a}\Psi_R) \approx 0,\\
& & \partial_{-}\widetilde{\Pi}^{-a}-g(\Pi^{-c}\epsilon^{abc}A^{b}_{-})_{\sim}
           -\sqrt{2}g(\Psi^{\dagger}_RT^{a}\Psi_R)_{\sim} \approx 0,\\
& & \sqrt{2}i\partial_{-}\Psi_L-m\Psi_R-\sqrt{2}g\Psi_LT^{a}A^{a}_{-}
\approx 0,\\
& & (\sqrt{2}i\partial_{-}\Psi_L-m\Psi_R-
            \sqrt{2}g\Psi_LT^{a}A^{a}_{-})^{\dagger} \approx 0,
\label{secondary}
\end{eqnarray}
where
\begin{equation}
(A_1A_2\cdot\cdot\cdot A_n)_{\sim} = A_1(x)A_2(x)\cdot\cdot\cdot A_n(x)
-\frac{1}{2L}\int_{-L}^{L}dx^{-}A_1(x)A_2(x)\cdot\cdot\cdot A_n(x).
\label{pmmode}
\end{equation}
We can show directly that constraints (2.34)-(2.37)  do not generate new
constraints further. Adding these constraints to the primary constraints,
 there exist ten constraints, which govern the dynamics of our
system.
 What we have to do next is to classify these primary and secondary constraints
 to the first class or the second class constraints.
A direct calculation shows that the constraints (2.26) and (2.27) belong to
the first
class and others the second one. But this is not true. As indicated by
some authors
\cite{sunder,gaete}, the minimal set of the second class constraints is found
by
combining constraints except for (2.26) and (2.27) appropriately and it is
easy to
show that this set is indeed given by
\begin{eqnarray}
\Omega^a_0 &=& \zm{\Pi^{+a}}, \\
\Omega^a_1 &=& \widetilde{\Pi}^{+a}(x), \\
\Omega^a_2 &=& g\int_{-L}^{L}dx^{-}[\Pi^{-c}\epsilon^{abc}A^b_{-}(x)
\nonumber \\
           & & \qquad \qquad  -i(P_LT^a\Psi_L+\Psi^{\dagger}_LT^aP^{\dagger}_L
               +P_RT^a\Psi_R+\Psi^{\dagger}_RT^aP^{\dagger}_R)(x)] , \\
\Omega^a_3 &=&
\left(\partial_{-}\Pi^{-a}(x)-g\Pi^{-c}\epsilon^{abc}A^{b}_{-}(x) \right.
\nonumber\\
           & & \left. \qquad \quad
+ig(P_LT^a\Psi_L+\Psi^{\dagger}_LT^aP^{\dagger}_L
               +P_RT^a\Psi_R+\Psi^{\dagger}_RT^aP^{\dagger}_R)(x)
\right)_{\sim},\\
\chi_1 &=& P_L(x), \\
\chi_2 &=& P^{\dagger}_L(x), \\
\chi_3 &=& P_R(x)-\sqrt{2}i\Psi^{\dagger}_R(x), \\
\chi_4 &=& P^{\dagger}_R(x), \\
\chi_5 &=&
\sqrt{2}i\partial_{-}\Psi_L(x)-m\Psi_R(x)-\sqrt{2}g\Psi_LT^{a}A^{a}_{-}(x),\\
\chi_6 &=& \left(\sqrt{2}i\partial_{-}\Psi_L(x)-m\Psi_R(x)-
           \sqrt{2}g\Psi_LT^{a}A^{a}_{-}(x)\right)^{\dagger},
\label{class}
\end{eqnarray}
where $\Omega^a_{\alpha}$($\alpha$ = 0,1,2,3) and $\chi_{\beta}$($\beta = 1
\sim 6$)
denote the first and second class constraints, respectively.
 The first class constraints satisfy the algebra
\begin{equation}
\{ \Omega^a_{\alpha}(x), \Omega^b_{\beta}(y) \} = 0,
\label{algebra}
\end{equation}
which reflects the gauge invariance of the system.
 In order to eliminate all the constraints and quantize the system, we need
to fix
the gauge degrees of freedom and define the Dirac bracket along the usual
prescriptions \cite{sunder}.
Here we shall give the gauge fixing conditions as follows:
\begin{eqnarray}
\omega^a_0 &\equiv& \zm{A^a_{+}} \approx 0, \\
\omega^a_1 &\equiv& \widetilde{\Pi}^{-a}(x)+\partial_{-}\widetilde{A}^a_{+}(x)
                +g\epsilon^{abc}\left(A^b_{+}(x)A^c_{-}(x)\right)_{\sim}
\approx 0,\\
\omega^a_2 &\equiv& \zm{A^i_{-}} \approx 0, \qquad for \qquad i = 1,2,\\
\omega^a_3 &\equiv&  \widetilde{A}^a_{-}(x) \approx 0.
\label{gaugefix}
\end{eqnarray}
Note here the following remarkable fact. As we see from the gauge-fixing
conditions
(2.52) and (2.53), we can not impose the traditional light-cone gauge
$A^a_{-}$ = 0
because we can not put the third component of $\zm{A^a_{-}}$ to be zero.
 $SU(2)$ global color rotation symmetry always enables us to choose such a
gauge
fixing conditions \cite{pinsky}.
 That is why one of the zero modes of the gluon field $\zm{A^3_{-}}$
becomes a dynamical variable, which
might give insight to the nontrivial structures of the QCD light-cone vacuum
\cite{pinsky}.

 Now we are coming in the stage of evaluating the Dirac bracket. After the
straightforward but some tedious calculations, non-zero Dirac brackets are
\begin{eqnarray}
\{ \zm{A}^3_{-}, \zm{\Pi^{-3}} \}_{DB} &=& \frac{1}{2L},\\
\{ \Psi_R(x),\Psi^{\dagger}_R(y) \}_{DB} &=&
\frac{i}{\sqrt{2}}\delta(x^{-}-y^{-}).
\label{diracbracket}
\end{eqnarray}
As far as Dirac brackets have been used, we may put all the constraints and
the gauge
fixing conditions to strongly zeroes.The result is that total Hamiltonian
(\ref{lchamiltonian1}) reduces to the form
\begin{equation}
H^{l.c.}_T = \int_{-L}^{L}dx^{-}\left[\frac{1}{2}p^2
             + m\Psi^{\dagger}_R\Psi_L
+\frac{g}{\sqrt{2}}(\Psi^{\dagger}_RT^a\Psi_R)_{\sim}
\widetilde{A}^a_{+}\right],
\label{lchamiltonian2}
\end{equation}
where $p \equiv \zm{\Pi^{-3}}$ and
$\Psi_L(x)$ and $\widetilde{A}^a_{+}(x)$ are given as the functions satisfying
following equations, i.e.,
\begin{eqnarray}
& &\sqrt{2}i\partial_{-}\Psi_L(x)-m\Psi_R(x)-\sqrt{2}gq(x^+)T^{3}\Psi_L(x)
= 0,\\
& &\partial^2_{-}\widetilde{A}^a_+(x)
             +2\epsilon^{ab3}gq(x^+)\partial_{-}\widetilde{A}^a_+(x)
\nonumber \\
      & & \qquad -g^2q^2(x^+)(\widetilde{A}^a_+
-\delta^{a,3}\widetilde{A}^3_+)(x)
            +\sqrt{2}g\Psi^{\dagger}_RT^a\Psi_R(x) = 0,
\label{psil}
\end{eqnarray}
where $q(x^+) \equiv \zm{A}^3_{-}$.
While $\Psi_R(x)$ has been chosen to satisfy charge neutrality condition
\begin{equation}
Q^3 \equiv g\int_{-L}^{L}dx^{-}\Psi^{\dagger}_RT^3\Psi_R(x) = 0.
\label{cneutrality}
\end{equation}
This corresponds to the third component of ZM of Gauss law,
which is necessary whenever the system
is in a finite interval \cite{landau}. The fields
$\Psi_L$ and $\widetilde{A}^a_+$
are expressed in terms of $\Psi_R$ and $q(x^+)$ by solving the equations (2.57)
and (2.58). As the result, we find that
the physical degrees of freedom in this system are
only the diagonal part of zero modes of the gluon field $q(x^+)$ and
right-handed
quark field $\Psi_R$.

\sect{Quantum Theory  - Dynamical ZM Equation - }

In this section, we discuss the quantum aspects of the theory studied in the
previous section at the classical level in detail.
First we start by discussing eigenstates of the matter part in the Hamiltonian
(\ref{lchamiltonian2}) in the fixed background gauge field and then we
construct
the full ground state including ZM of the gauge field $q(x^+)$. Before
doing that,
we must solve equations (2.57) and (\ref{psil}) for $\Psi_L$ and
$\widetilde{A}^a_+$. Equation (2.57) for $\Psi_L$ is easy to be solved as
follows:
\begin{equation}
\Psi^{c}_L(x) = \frac{m}{2\sqrt{2}L}\sum_{k=0}^{\infty}\int_{-L}^{L}dy^{-}
                 e^{-\frac{i\pi}{L}(k+\frac{1}{2})(x^{-}-y^{-})}
                  \widetilde{\Psi}^{c}_{R}(y;k),
\label{psil2}
\end{equation}
where
\begin{equation}
\widetilde{\Psi}^{c}_{R}(x;k) = \left( \begin{array}{c}
    \frac{\Psi^1_R(x)}{\frac{\pi}{L}(k+\frac{1}{2})-gq} \\
    \frac{\Psi^2_R(x)}{\frac{\pi}{L}(k+\frac{1}{2})+gq}
                                  \end{array} \right)^c,
\label{psilR}
\end{equation}
with $c$ being the color indices.

 On the other hand, equations for $\widetilde{A}^a_+$ consist of the following
three components:
\begin{eqnarray}
& & \partial^2_{-}\widetilde{A}^1_+(x) +2gq(x^+)\partial_{-}\widetilde{A}^2_+
              -g^2q^2(x^+)\widetilde{A}^1_+ + \rho^1(x) = 0, \\
& & \partial^2_{-}\widetilde{A}^2_+(x) -2gq(x^+)\partial_{-}\widetilde{A}^1_+
              -g^2q^2(x^+)\widetilde{A}^2_+ + \rho^2(x) = 0, \\
& & \partial^2_{-}\widetilde{A}^3_+(x)  + \rho^3(x) = 0,
\label{widetidea}
\end{eqnarray}
where $\rho^a(x) \equiv \sqrt{2}g\Psi^{\dagger}_RT^a\Psi_R(x)$. Clearly a
solution
for  equation (3.5) is formally written of the form $\widetilde{A}^3_+(x) =
- \frac{1}{\partial^2_{-}}\rho^3(x)$. Thus, we will concentrate to the remained
equations (3.3) and (3.4).
 For brevity, we rewrite equations (3.3) and (3.4)  as
\begin{eqnarray}
\frac{d^2f(x)}{dx^2} +2a\frac{dg(x)}{dx} -a^2f(x) + \rho^1(x) =0, \nonumber \\
\frac{d^2g(x)}{dx^2} -2a\frac{df(x)}{dx} -a^2g(x) + \rho^2(x) =0,
\label{diffeq}
\end{eqnarray}
where we are putting now as follows:
\begin{eqnarray}
x &\equiv& x^-,\nonumber \\
 a &\equiv& gq(x^+),\nonumber \\
 \widetilde{A}^1_+(x) &\equiv& f(x),\nonumber \\
 \widetilde{A}^2_+(x) &\equiv& g(x).
\label{simple}
\end{eqnarray}
It is easy to find that we can express equation (\ref{diffeq}) in matrix
representation
\begin{equation}
M^{ij}(x)\varphi^{j}(x) = -\rho^i(x).
\label{matrix1}
\end{equation}
Here 2 $\times$ 2 matrix $M^{ij}(x)$ and vectors $\phi^{i}(x)$ and
$\rho^{i}(x)$
are defined by
\begin{eqnarray}
M^{ij}(x) &=& \left(\begin{array}{cc}
            \frac{d^2}{dx^2}-a^2 & 2a\frac{d}{dx} \\
            -2a\frac{d}{dx} & \frac{d^2}{dx^2}-a^2 \end{array}\right)^{ij},
           \nonumber \\
 \varphi^i(x) &=& \left( \begin{array}{c}
               f(x) \\ g(x)
                \end{array} \right)^i,
\qquad \rho^i(x) = \left( \begin{array}{c}
               \rho^1(x) \\ \rho^2(x)
                \end{array} \right)^i.
\label{matrix2}
\end{eqnarray}
Using the usual Green function's method, the form of $\varphi^{i}(x)$
would be given by
\begin{equation}
\varphi^{i}(x) = \int_{-L}^{L}dyG^{ij}(x,y)\rho^j(y) ,
\label{solution}
\end{equation}
where  $G^{ij}(x,y)$ is the Green function defined by
\begin{equation}
M^{ij}(x)G^{jk}(x,y) = -\delta^{ik}\delta(x-y).
\label{greenfunc}
\end{equation}
By solving equation (\ref{greenfunc}) with $M^{ij}(x)$ given by
(\ref{matrix2}),
we obtain the explicit forms of $G^{ij}(x,y)$, $\widetilde{A}^1_{+}(x)$ and
$\widetilde{A}^2_{+}(x)$ such that
\begin{eqnarray}
G^{ij}(x,y) &=& \frac{1}{L}\sum_{n=-\infty}^{\infty}
            \frac{e^{\frac{\pi ni}{L}(x-y)}}
            {(\frac{\pi n}{L}+a)^2(\frac{\pi n}{L}-a)^2}
             \left(\begin{array}{cc}
            (\frac{\pi n}{L})^2+a^2 & \frac{2\pi ina}{L} \\
            -\frac{2\pi ina}{L} & (\frac{\pi n}{L})^2+a^2
\end{array}\right)^{ij},\\
\label{greenfunc2}
\nonumber \\
\widetilde{A}^1_{+}(x) &=& \frac{1}{L}\sum_{n=-\infty}^{\infty}
                        \int_{-L}^{L}dy
                      \frac{[(\frac{\pi n}{L})^2+a^2]\rho^1(y)
                       +\frac{2\pi ina}{L}\rho^2(y)}
                      {(\frac{\pi n}{L}+a)^2(\frac{\pi n}{L}-a)^2}
                       e^{\frac{\pi ni}{L}(x-y)}, \\
\nonumber \\
\widetilde{A}^2_{+}(x) &=& \frac{1}{L}\sum_{n=-\infty}^{\infty}\int_{-L}^{L}dy
                      \frac{[(\frac{\pi n}{L})^2+a^2]\rho^2(y)
-\frac{2\pi ina}{L}\rho^1(y)}
{(\frac{\pi n}{L}+a)^2(\frac{\pi n}{L}-a)^2}e^{\frac{\pi ni}{L}(x-y)}.
\label{aparticle}
\end{eqnarray}
As the result, by substituting these equations and $\widetilde{A}^3_+(x) =
- \frac{1}{\partial^2_{-}}\rho^3(x)$ into the electrostatic Coulomb energy part
in the Hamiltonian, then we find that
\begin{eqnarray}
H^{l.c.}_{Coulomb} &\equiv& \int_{-L}^{L}dx^{-}
\frac{g}{\sqrt{2}}(\Psi^{\dagger}_RT^a\Psi_R)_{\sim}\widetilde{A}^a_{+}(x),
\nonumber \\
&=& \frac{1}{2}\int_{-L}^{L}dx^{-}[\rho^1(x)\widetilde{A}^1_{+}
+ \rho^2(x)\widetilde{A}^2_{+} + \rho^3(x)\widetilde{A}^3_{+}], \nonumber \\
& & -\frac{1}{2}\int_{-L}^{L}dx^{-}\rho^3(x)\frac{1}{\partial^2_{-}}\rho^3(x),
\nonumber \\
&=& \int_{-L}^{L}dx^{-}\int_{-L}^{L}dy^{-}\sum_{i=1,2}\rho^i(x)\rho^i(y)
e^{-igq(x^+)T^3(x^{-}-y^{-})} \nonumber \\
& & \times \left( \frac{L}{2\sin^2(gqL)}
+i(x^{-}-y^{-})\cot(gqL) - |x^{-}-y^{-}| \right) \nonumber \\
& &
-\frac{1}{2}\int_{-L}^{L}dx^{-}\rho^3(x)\frac{1}{\partial^2_{-}}\rho^3(x).
\label{coulomb}
\end{eqnarray}
This is consistent with the result from \cite{wadia}.

 In order to quantize the Hamiltonian (\ref{lchamiltonian2}),
we replace the Dirac brackets to the commutators.
Quantization conditions for the fields $\Psi_R(x)$ and $q(x^+)$ are defined as
\begin{eqnarray}
[\widehat{q}(x^+), \widehat{p}(x^+)] &=& i, \\
\{ \widehat{\Psi}^c_R(x), \widehat{\Psi}^{c'\dagger}_R(y) \}
               &=& \frac{1}{\sqrt{2}}\delta^{c,c'}\delta(x^{-}-y^{-}), \\
\{ \widehat{\Psi}^c_R(x), \widehat{\Psi}^{c'}_R(y) \} &=&
\{ \widehat{\Psi}^{c\dagger}_R(x), \widehat{\Psi}^{c'\dagger}_R(y) \} =
0,                 \label{commutators}
\end{eqnarray}
where $\widehat{A}$ means an operator and we have rescaled $2Lq \to q$.
 Thus, quantum light-cone Hamiltonian is composed of following three parts:
\begin{equation}
\widehat{H}^{l.c.} = \widehat{H}^{l.c.}_{ZM} +\widehat{H}^{l.c.}_{F}
+ \widehat{H}^{l.c.}_{Coulomb},
\label{qhamiltonian}
\end{equation}
where
\begin{eqnarray}
\widehat{H}^{l.c.}_{ZM} &=& \int_{-L}^{L}dx^{-}\frac{1}{2}\widehat{p}^2, \\
\widehat{H}^{l.c.}_{F} &=& \int_{-L}^{L}dx^{-}m\widehat{\Psi}_R^{\dagger}
\widehat{\Psi}_L, \\
\widehat{H}^{l.c.}_{Coulomb} &=&
\int_{-L}^{L}dx^{-}\int_{-L}^{L}dy^{-}\sum_{i=1,2}
\widehat{\rho}^i(x)\widehat{\rho}^i(y)
e^{-\frac{ig}{2L}\widehat{q}(x^+)(x^{-}-y^{-})} \nonumber \\
& & \times \left( \frac{L}{2\sin^2(\frac{g\widehat{q}}{2})}
+i(x^{-}-y^{-})\cot(\frac{g\widehat{q}}{2}) - |x^{-}-y^{-}| \right) \nonumber
\\
& & - \frac{1}{2}\int_{-L}^{L}dx^{-}\widehat{\rho}^3(x)
\frac{1}{\partial^2_{-}}\widehat{\rho}^3(x),
\label{commutators2}
\end{eqnarray}
and
\begin{eqnarray}
\widehat{\Psi}^{c}_L(x) &=& \frac{m}{2\sqrt{2}L}\sum_{k=0}^{\infty}
           \int_{-L}^{L}dy^{-}e^{-\frac{i\pi}{L}(k+\frac{1}{2})(x^{-}-y^{-})}
                  \widehat{\widetilde{\Psi}}^{c}_{Rk}(y), \\
\widehat{\rho}^a(x) &=&
\sqrt{2}g:\widehat{\Psi}^{\dagger}_RT^a\widehat{\Psi}_R(x):.
\label{current}
\end{eqnarray}
Here : : means a normal-ordered product.

 In this stage, our treatment is still exact.
Since we could obtain the complete forms of quantum light-cone Hamiltonian
$\widehat{H}^{l.c.}$, we will next discuss eigenstates of $\widehat{H}^{l.c.}_F
+\widehat{H}^{l.c.}_{Coulomb}$. To do that, we first construct the ground
state of
our system in the presence of fixed background values of
ZM of the gauge field $q(x^+)$ \cite{wadia}.
This is corresponding to a kind of the adiabatic approximation.
Then we can find easily a fermionic part of the light cone vacuum
eigenstate as
$|0\rangle_f$, defined by
\begin{eqnarray}
\widehat{a}^c_k|0\rangle_f = \widehat{d}^c_k|0\rangle_f = 0, \nonumber
\label{fermivac}
\end{eqnarray}
for all $c$ and $k$. Here the creation (annihilation) operator $\widehat{a}^{c
\dagger}_k,\widehat{d}^{c \dagger}_k (\widehat{a}^c_k,\widehat{d}^c_k)$
 are defined through the following mode expansion of
$\widehat{\Psi}_R(x)$, which comes from the antisymmetric boundary
condition (2.9)
and the anticommutation relation (3.17) and (3.18):
\begin{equation}
\widehat{\Psi}^{c}_R(x) = \frac{1}{2^{\frac{1}{4}}\sqrt{2L}}\sum_{k=0}^{\infty}
 \{ \widehat{a}^{c}_{k}e^{-\frac{i\pi}{L}(k+\frac{1}{2})x^-}
   +\widehat{d}^{c\dagger}_{k}e^{\frac{i\pi}{L}(k+\frac{1}{2})x^-} \},
\label{modeexp}
\end{equation}
where
\begin{eqnarray}
\{ \widehat{a}^{c}_{k}, \widehat{a}^{c'\dagger}_{k'} \}
= \delta^{c,c'}\delta_{k,k'} =
\{ \widehat{d}^{c}_{k}, \widehat{d}^{c'\dagger}_{k'} \}, \nonumber \\
\{ \widehat{a}^{c}_{k}, \widehat{a}^{c'}_{k'} \}
= \{ \widehat{a}^{c\dagger}_{k}, \widehat{a}^{c'\dagger}_{k'} \} = 0,
\nonumber \\
\{ \widehat{d}^{c}_{k}, \widehat{d}^{c'}_{k'} \}
= \{ \widehat{d}^{c\dagger}_{k}, \widehat{d}^{c'\dagger}_{k'} \} = 0.
\label{creannihi}
\end{eqnarray}
Of course, we can easily show that the states $|0\rangle_f$ satisfies the
constraints (\ref{cneutrality}), that is,
\begin{equation}
\widehat{Q}^3|0\rangle_f \equiv g\int_{-L}^{L}dx^{-}
\widehat{\Psi}^{\dagger}_R(x)T^3\widehat{\Psi}_R(x)|0\rangle_f = 0.
\label{physical}
\end{equation}
This is nothing but the physical state condition. In physical meaning, this
is saying
that physical states be charge neutral as a whole.

 As the result, the full ground state of the system can be written by
\begin{equation}
|vacuum\rangle  \cong |0\rangle_f \otimes \Phi_0(q).
\label{fullvac}
\end{equation}
$\Phi_0(q)$ is the zero mode wave function in the $q$ representation,
satisfying the following Schr\"{o}dinger equation for a free particle
with a unit mass ($m=1$)
\begin{equation}
\frac{1}{2} \left(-i\frac{d}{dq}\right)^2\Phi_0(q) = {\cal E}\Phi_0(q),
\label{schrod}
\end{equation}
where ${\cal E} \equiv E/(2L)$ is an energy density.

 This result seems to suggest that in a sense, the ground state structure of
the
correctly normal-ordered light-cone QCD Hamiltonian is almost trivial in
the adiabatic
approximation. However, it is difficult to construct the full ground state
beyond the adiabatic approximation. Rather, we are interested in how the
effects
of ZM change the spectrum of the excited states. But we can not answer this
question
in this paper.

 Instead of it, we shall see the relation between our result and that
 of Kalloniatis et al.. In order to do so, we shall neglect the fermion
mass term
 (3.21) and replace the currents $\widehat{\rho}^i(x) (i=1,2)$ in eq.(3.22)
with classical external source terms $\rho^i$ independent of $x$
(note that $\rho^3$ automatically vanishes
because of the charge neutrality condition (2.59)).
They have assumed in their paper that only zero mode external sources
excite ZM of the gauge fields.
After some straightforward calculations of $x-$ and $y-$ integrations,
we find that the light-cone Hamiltonian
in the $q$ representation would be of the form as
\begin{eqnarray}
H^{l.c.} &=& H^{l.c.}_{ZM} \\ \nonumber
                   &+&4L\left[ (\rho^1L)^2+(\rho^2L)^2 \right] \\ \nonumber
             &\times& \left[\frac{1}{gq}
              \left( \cot(\frac{gq}{2})
             -2\cot(\frac{gq}{2})\cos(gq)
             -4\sin(gq) \right)
              +\frac{4}{g^2q^2}\cos^2(\frac{gq}{2}) \right].
\label{kalloh}
\end{eqnarray}
Moreover in a specific coupling region, i.e., weak coupling region
($gq \ll 1$), the above equation reduces to
\begin{eqnarray}
H^{l.c.} &=& H^{l.c.}_{ZM}
             + 8L\left[ (\rho^1L)^2+(\rho^2L)^2 \right]\frac{1}{g^2q^2} \\
\nonumber
          &=& 2L\left[\frac{1}{2}\left(-i\frac{d}{dq}\right)^2
               +\frac{(2wL)^2}{2q^2} \right],
\label{kalloh2}
\end{eqnarray}
where
\begin{equation}
w^2 \equiv \frac{\rho_+\rho_-}{g^2},
\qquad \qquad \rho_{\pm} \equiv \sqrt{2}(\rho^1 \pm i\rho^2).
\label{redef}
\end{equation}
Therefore Schr\"{o}dinger equation for dynamical zero mode is now given by
\begin{equation}
     \frac{1}{2}\left[-\frac{d^2}{dq^2}+\frac{(2wL)^2}{q^2} \right]\Phi_0(q)
       = {\cal E}\Phi_0(q).
\label{schro2}
\end{equation}
This is the same result Kalloniatis et al. have obtained in ref.
\cite{kallo}. They have also used a kind of weak coupling approximation
to obtain their result.
\sect{Summary and Discussion}

 In this paper, we have studied $QCD_{1+1}$ with fundamental fermions
based on Discretized Light-Cone Quantization (DLCQ) formalism.
We have discussed both classical and quantum aspects of the theory in detail
and obtained the full ground state wave function of the theory by the method of
\lq \lq separation of variables" mentioned in ref.\cite{wadia}
and we could see the light-cone QCD ground state has almost trivial structure
in the range of the adiabatic approximation we have used here.
Also we could find the relation between the original work by Kalloniatis et al.
and ours.
 The physical effects of ZM for the essentially nonperturbative
phenomena e.g. chiral symmetry breaking and
confinement etc., however, remains unclear.
More precise considerations for these would be the future work.
 Moreover, what we would like to really understand is {\it QCD bound-state
problem
in 3+1 dimensions} \cite{wilson}. But as there are many difficulties
to get there, especially renormalization problem \cite{perry2}, it seems far
long way. Future works will be concentrated on these points.


\vspace{1cm}
\begin{center}
{\Large\bf Acknowledgments \par}
\end{center}

We would like to thank K. Harada, M. Taniguchi, T. Sugihara, A. Okazaki and
S. Sakoda for helpful discussions in Kyushu University.
We also appreciate  M. Sakamoto, H. Sato and K. Maeda for their kindness.

\newpage
\section*{Appendix: Notations and Conventions}

 We describe here some notations and conventions in light-cone formalism.
They are essentially the same as those by Harada et al \cite{harada}.
the coordinates are set $x^{\pm}= (x^0 \pm x^1)/ \sqrt{2}$, where $x^+$ is
taken
as \lq \lq time". The light-cone metric is given by
\begin{equation}
g_{\mu \nu} = \left(\begin{array}{cc}
            0 & 1 \\
            1 & 0 \end{array}\right) = g^{\mu \nu}.  \qquad \mu,\nu= +,-
\nonumber
\end{equation}
The derivatives are also defined as
\begin{equation}
\partial_{\pm} \equiv \frac{\partial}{\partial x^{\pm}},
 \nonumber
\end{equation}
with $\partial_{\pm}=\partial^{\mp}$. $\gamma$ matrices in a representation
in which $\gamma^5$ is diagonal are as follows:
\begin{eqnarray}
\gamma^0 &=& \left(\begin{array}{cc}
             0 & 1 \\
             1 & 0 \end{array}\right),
\gamma^1 =\left(\begin{array}{cc}
            0 & -1 \\
            1 & 0 \end{array}\right),
\gamma^5 \equiv \gamma^0\gamma^1 =\left(\begin{array}{cc}
            1 & 0 \\
            0 & -1 \end{array}\right),\nonumber \\
\gamma^+  &=& \left(\begin{array}{cc}
            0 & 0 \\
            \sqrt{2} & 0 \end{array}\right),
\gamma^-  =\left(\begin{array}{cc}
            0 & \sqrt{2} \\
            0 & 0 \end{array}\right), \\ \nonumber
\gamma^+\gamma^-  &=& \left(\begin{array}{cc}
                 0 & 0 \\
                 0 & 2 \end{array}\right),
\gamma^- \gamma^+  =\left(\begin{array}{cc}
                  2 & 0 \\
                  0 & 0 \end{array}\right).
\end{eqnarray}
The SU(2) gauge fields are represented by
\begin{equation}
A_{\mu} \equiv A^{a}_{\mu}T^a, \quad T^a = \frac{1}{2}\sigma^a,
\qquad a=1,2,3 \nonumber
\end{equation}
where $\sigma^a$ is ordinary Pauli matrices such that
\begin{equation}
\sigma^1 = \left(\begin{array}{cc}
             0 & 1 \\
             1 & 0 \end{array}\right),
\sigma^2 = \left(\begin{array}{cc}
             0 & -i \\
             i & 0 \end{array}\right),
\sigma^3 = \left(\begin{array}{cc}
             1 & 0 \\
             0 & -1 \end{array}\right).
\end{equation}
%
%


\end{document}